\documentclass[10pt, journal]{IEEEtran}

\usepackage{amsmath, amsfonts}
\usepackage{mathrsfs}
\usepackage{times}
\usepackage{etoolbox}
\usepackage{romannum}
\usepackage{algorithm}
\usepackage{algpseudocode}
\usepackage{array}
\usepackage[caption=false, font=normalsize, labelfont=sf, textfont=sf]{subfig}
\usepackage{textcomp}
\usepackage{stfloats}
\usepackage{url}
\usepackage{verbatim}
\usepackage{graphicx}
\usepackage[numbers, sort&compress]{natbib}

\begin{document}

\title{CorrNetDroid: Android Malware Detector leveraging a Correlation-based
Feature Selection \\ for Network Traffic features}

\author{Yash Sharma and Anshul Arora
\thanks{ysharma2098@gmail.com and anshul15arora@gmail.com.\\Department of Applied Mathematics, Delhi Technological University, Delhi -110042, India}}




\maketitle

\begin{abstract}
Copious mobile operating systems exist in the market, but the Android remains the user’s choice. Meanwhile, its growing popularity has also allured malware developers. Researchers have proposed various static solutions for Android malware detection. However, stealthier malware evade static analysis. This raises the need for a robust Android malware detection system capable of dealing with advanced threats and overcoming the shortcomings of static analysis. Hence, in this work, we have proposed a dynamic analysis-based Android malware detection system, \textit{CorrNetDroid}, that works over network traffic flows. Many of the traffic features have overlapping ranges in normal and malware datasets. Therefore, we first rank the features using two statistical measures namely, \textit{crRelevance} and \textit{Normalized Mean Residue Similarity}, assessing feature–class and feature–feature correlations. Thereafter, we propose a novel correlation-based feature selection algorithm that deploys \textit{NMRS} on the \textit{crRelevance} rankings to identify the optimal feature subset for Android malware detection.
The experimental results highlight that our proposed model can effectively reduce the feature set while detecting Android malware with 99.50\% accuracy on considering only two network traffic features. Furthermore, our experiments demonstrate that the \textit{NMRS}-based proposed algorithm on \textit{crRelevance} rankings is better than other statistical tests such as chi-square, ANOVA, Mann–Whitney U test and Kruskal–Wallis test. In addition, our proposed model outperforms various state-of-the-art Android malware detection techniques in terms of detection accuracy. 
\end{abstract}


\section{Introduction}

What started as a mere device for communication purposes has now become the smartphone era. We have gradually accepted smartphones as an indispensable part of our lives because of the need for communication and other daily necessities. These days, one can find an application for almost everything. Statistics depict the emergence of Android apps, as over the past 14 years, the Google Play Store has expanded significantly, increasing from 16,000 apps in 2009 to 3.553 million apps. \footnote{https://www.bankmycell.com/blog/number-of-google-play-store-apps/)}. As for the most preferred operating system, Android OS reigns at the top due to its numerous attributes. Open architecture and the availability of various feature-rich applications are just some of the merits of Android OS, and hence, Android mobile sales have grown exponentially in the past 14 years. \\



With its increasing popularity worldwide, Android has become a victim of its own success as malware authors constantly strive to steal private information or even perpetrate advanced fraud by hiding malicious code inside applications without the user’s consent. According to Kaspersky Security Network, in the first quarter of 2024, close to 10 million mobile malware incursions were blocked \footnote{https://securelist.com/it-threat-evolution-q1-2024-mobile-statistics}. To address these troubling figures, researchers have suggested a range of malware detection methods in the literature. Generally, the detection approaches proposed in the literature can be classified into two categories, namely static, and dynamic. \cite{sharma2024comprehensive}.


Static analysis has demonstrated significant efficiency in both feature extraction and cost-effectiveness. Examining malware without running the code itself and gathering fundamental details about an application's behavior appeared to be advantageous. 
However, some stealthier malware may not show their malicious behavior in static analysis, i.e., malware with advanced techniques such as code obfuscation, polymorphism, and encryption may evade static detection. Dynamic analysis, wherein applications are monitored at runtime, overcomes the shortcomings of static analysis. Therefore, in this study, we aim to develop a dynamic malware detection framework.

Stealthier malware samples establish connections with remote servers clandestinely to receive commands or to exfiltrate users' private information or device data to the server. 
Approximately 93\% of Android malware exhibit network connectivity. Since these  malware are remotely controlled, they effectively transform the mobile device into a mobile bot, presenting a significant threat to the user community. 

Some dynamic Android malware detection techniques have been proposed in the literature using features such as system calls \cite{singh2017dynamic}, cryptographic and network operation \cite{feng2018novel}. However, we chose network traffic flows  to perform dynamic analysis because they provide a comprehensive and real-time view of an application’s external communication. 
Simultaneously, features such as system calls provide limited visibility into external communications, especially if the malware disguises its activities by leveraging legitimate APIs or system calls. Moreover, anomalies may be harder to detect because some malware can mimic legitimate application behavior and alter its API or system call patterns to avoid detection. Hence we aim to analyze network traffic features for Android malware detection.


\begin{table}[b]
\fontsize{7}{9}\selectfont
\centering
\caption{ Range of traffic features for malware and normal mobile applications }{\label{Range_traffic_features}}

\begin{tabular}{|p{1.25in}|p{.8in}|p{.8in}|} \hline

\textbf{Feature Name }                & \textbf{Range in normal traffic} & \textbf{Range in malware traffic }\\ \hline
Flow\_duration               & 0 - 49560.3403         & 0 - 52122.128886        \\ \hline
Packets\_sent\_per\_flow     & 1 - 110985             & 1 - 48265               \\ \hline
Packets\_received\_ per\_flow & 0 - 215014             & 0 - 104312              \\ \hline
Bytes\_sent                  & 40 - 82771640          & 54 - 16288912           \\ \hline
Bytes\_received              & 0 - 1217408141         & 0 - 157283100           \\ \hline
\end{tabular}
\end{table}

\subsection{Motivation} 

Analyzing network traffic usage patterns is an effective way to detect malware. As a result, network traffic flows have been extensively utilized in Android malware detection. Nonetheless, there exist numerous resemblances in the traffic feature patterns between benign and malicious applications. Tables \ref{Range_traffic_features} summarizes the range of several traffic features. We extracted over nine lacs network traffic flows of the normal class and an equal number for the malware class by combining the datasets from various repositories. Additional information regarding the dataset is presented in subsequent sections. Furthermore, we extracted 16 network traffic features from the network traffic data.  As seen in Table \ref{Range_traffic_features}, commonly used features, namely ``Flow\_Duration'', ``Packets\_sent\slash received'', ``Bytes\_sent\slash received'' are present in overlapping ranges when observed in both normal and malware traffic.

The presence of these similar features in both datasets serves as a motivation for us to prioritize feature ranking. The aim of our proposed work is to present an efficient detection model that utilizes prioritized traffic features. 



Several related studies have used the dynamic traffic features  for Android malware detection. For instance, Wang et al. \cite{wang2019mobile} analyzed multiple levels of network traffic features and emphasized that combining 2 levels, namely HTTP packet and TCP flow, can successfully lead to malware detection. 
However, they failed to consider the crucial idea of feature ranking, hence neglecting the step of feature reduction, which had the potential to enhance the quality of their results.

In several other related works, such as \cite{arora2017minimizing} and \cite{shabtai2014mobile}, the authors built a detection system using the best subset of features by feature ranking. More specifically, Arora and Peddoju \cite{arora2017minimizing} extracted 22 network traffic features. Consequently, they aimed to reduce the feature set and used information gain and chi-square test to rank the feature set. 
In \cite{shabtai2014mobile}, the authors tried to understand the reason behind the deviations in the application’s network traffic behavior from the normal flow by observing the network traffic flows. Hence, they computed the probability scores depicting the deviation of features’ behaviors from normal traffic patterns and later used the threshold approach to select the best subset of features. Nevertheless, both studies were carried out with a limited number of network traffic flows in comparison to the comprehensive dataset employed in our proposed study. More importantly, our research demonstrates superior performance as compared to both in terms of detection accuracy.
 

Our experiments demonstrate that our proposed \textit{NMRS}-based detection model with \textit{crRelevance} rankings outperforms other statistical tests such as chi-square, ANOVA, Mann– Whitney U test, Kruskal–Wallis test. 

\subsection{Research Objectives}

The proposed dynamic detection approach centering the ranking of network traffic attributes gives rise to the following research questions:

\begin{enumerate}

\item Why is there a need to prioritize or rank network traffic features?

\item How to incorporate feature ranking while eliminating redundant features, i.e., how to rank network traffic features and select the least correlated subset?

\item How to frame a detection approach while considering both feature– class and feature– feature correlations?

\item What is the advantage(s) of feature ranking regarding detection accuracy?

\end{enumerate}
\textbf{Organization}: 
In Section II, we examine some previous research in the field. This is followed by a thorough description of the proposed approach in Section III. Hereafter, we provide the results acquired using the proposed model in Section IV along with addressing certain limitations. Lastly, we wrap up the work by discussing potential areas for future research in Section V.

\section{Related Work}
 We deliver a comprehensive analysis of existing research in Android malware detection in this section. We focus our discussion on dynamic detection models. Researchers have used dynamic features such as function call graphs, system calls \cite{singh2017dynamic}, cryptographic and network operation \cite{feng2018novel} for Android malware detection. However, we chose network traffic flows because they provide a comprehensive and real-time view of an application’s external communication, making it particularly effective in the context of Android malware detection. Simultaneously, system calls dispense limited visibility into external communications, especially if the malware disguises its activities by leveraging legitimate APIs or system calls. Besides, anomalies may be harder to detect because some malware can mimic legitimate application behavior and alter its API or system call patterns to avoid detection. Cryptographic operations focus on encryption and decryption routines but may miss the extensive context of malicious activities. Hence, we use dynamic network traffic features in this study.

In this section, we aim to discuss studies that have used TCP, HTTP, or DNS flows for Android malware detection. We divide the related work into two categories: network traffic-based analysis and traffic-based detection. 

 \subsection{Network Traffic-Based Analysis}

In the literature, few studies have aimed to analyze network traffic of mobile applications \cite{analyzing1, analyzing2}. These works did not consider malware apps. Some other techniques have analyzed the dynamic features of TCP flows, HTTP requests, and DNS queries concerning malicious Android apps. In this subsection, we review all of these techniques.

The authors in \cite{chen_a_first_look} analyzed the network traffic of malware samples during the first five minutes of installation and operation. They concluded that the majority of the network traffic is covered by HTTP and DNS flows. The authors in \cite{aresu2015clustering} analyzed the HTTP traffic generated by mobile botnets to group their families effectively. They observed statistical information related to HTTP traffic and formed malware clusters.  Kumar and Sharma \cite{Understanding_Ransom_2023} focused on analyzing the Android Ransomware network architecture and suggested a road map for detecting such attacks. All of these studies analyzed the malware traffic and did not aim for malware detection.


\subsection{Network Traffic-Based Detection}

In the literature, several authors have focused on Android malware detection using network traffic features. This section provides a comprehensive review of all such related works proposed in the literature. We further divide this section into two subsections: network traffic-based detection with feature ranking and traffic-based detection without feature ranking.

\subsubsection{Detection with Feature Ranking} 

First, we examine methods that have utilized any feature ranking\slash selection technique for Android malware detection.

Arora and Peddoju \cite{arora2018ntpdroid} proposed a hybrid detection model to combine the permissions and network traffic features extracted from Android applications. They created a hybrid feature vector and applied the FP-Growth algorithm to extract frequent patterns in combinations of permissions and network traffic features for the benign and malware datasets. In their other work \cite{arora2017minimizing}, they extracted 22 network traffic features and further reduced the feature set with information gain and chi-square test. Upadhayay et al. \cite{upadhayay2021rpndroid} ranked the permissions based on their frequency in the benign and malware dataset. Next, to enhance their results, they introduced a dynamic analysis approach and coalesced the best set of network traffic features from their previous research work \cite{arora2017minimizing} to the newly obtained best set of permissions for malware detection.

The authors in \cite{wang2017detecting} proposed an approach analyzing the text semantics of HTTP request headers with the natural language processing algorithm called N-gram. They reduced the feature set to the most influential ones by applying the statistical chi-square test. Hossain et al. \cite{Hossain_2022} aimed to find a robust solution for growing ransomware attacks using network traffic features. They exploited particle swarm optimization to select only the optimal traffic characteristics and applied machine learning classifiers for malware detection.  The authors in \cite{lu2022f2dc} devised a deep learning-based malware classification system using raw payload and CNNs. They concentrated on the raw payload of malicious network traffic, leveraging the byte data to represent related behavioral patterns of malware, and further treated the window selection algorithm’s filtered (SWS) flows as documents to be processed by NLP methods.

To the best of our knowledge, no other study has assessed the ranking of dynamic attributes of TCP flows using a statistical \textit{crRelevance} test to detect Android malware. We chose \textit{crRelevance} for its simplicity and strong alignment with the goals of our research. Unlike other statistical tests, which often require strict assumptions such as normality of data, homogeneity of variances, and the independence of observations within mutually exclusive categories, \textit{crRelevance} is free from these constraints. These benefits make \textit{crRelevance} the most appropriate choice for our analysis. In this study, we proposed an \textit{NMRS}-based feature selection algorithm with \textit{crRelevance} rankings to detect Android malware with the most optimal attributes.

\subsubsection{Detection without Feature Ranking }

Within the literature, several detection mechanisms exist for detecting Android malware without any feature ranking\slash selection technique. This indicates that they have not given priority to the ranking of features. Within this subsection, we present a thorough review of these detection approaches.

 Shabtai et al. \cite{shabtai2014mobile} analyzed the reason behind the deviations in the application’s network traffic behavior. The authors focused on the server side of the system and believed that a strong correlation exists between benign behavior patterns that can be used to detect abnormal activities. Pang et al. \cite{pang2017finding} developed an efficient and convenient network traffic collection system. The authors observed that the malware applications produced a negligible amount of network traffic data; hence, they combined imbalance algorithms with the classical machine learning algorithms for their testing methods. The authors in \cite {wang2019mobile} analyzed multiple levels of traffic features and emphasized that combining two levels, namely HTTP packet and TCP flow, can be successfully employed to create a lightweight server-based malware detection model. 

Mahdavifar et al. \cite{Mahdavifar_Dynamic_Android} emphasized the concern regarding the difficulty in accumulating well-labeled datasets and hence proposed a Deep Neural Network (DNN) based work to perform dynamic analysis for malware category classification. 
Liu et al. \cite{liu2023nt} suggested a Graphical Neural Network (GNN) model based on mobile network traffic with node characteristics and edge attributes. The authors in \cite{alshehri2022app}  focused on repackaged apps. They computed the flow similarity using the Euclidian algorithm between two sets of network traffic features. 

Ullah et al. \cite{ullah2023nmal} proposed a multi-headed ensemble neural network by combining a convolutional neural network and gated recurrent units using the semantic features of HTTP traces and the numeric values of TCP flows. The authors in \cite{fallah2022android} employed a sequential-based deep learning (DL) model and stimulated network traffic data as a sequence of flows built upon the concept of long short-term memory. Similar to the previous work, the authors in \cite{shen2023self} used a deep learning-based model to detect malicious Android applications but used the network traffic flows as grayscale images instead of their original numeric form. 

We have highlighted the importance of ranking network traffic features for effective Android malware detection in the Introduction section. In addition, none of the above-discussed techniques applied feature ranking techniques with dynamic traffic features for mobile malware detection. In this study, we first ranked the network traffic features with \textit{crRelevance} to find the traffic feature set with the best ability to distinguish between benign and malicious class labels. Furthermore, we proposed an \textit{NMRS} based feature selection algorithm with \textit{crRlevance} to identify Android malware using the most effective features. We outline our proposed methodology in the following section.

\section{System Design}

\begin{figure}[t]
  
  \centering
  \includegraphics[width=.40\textwidth]{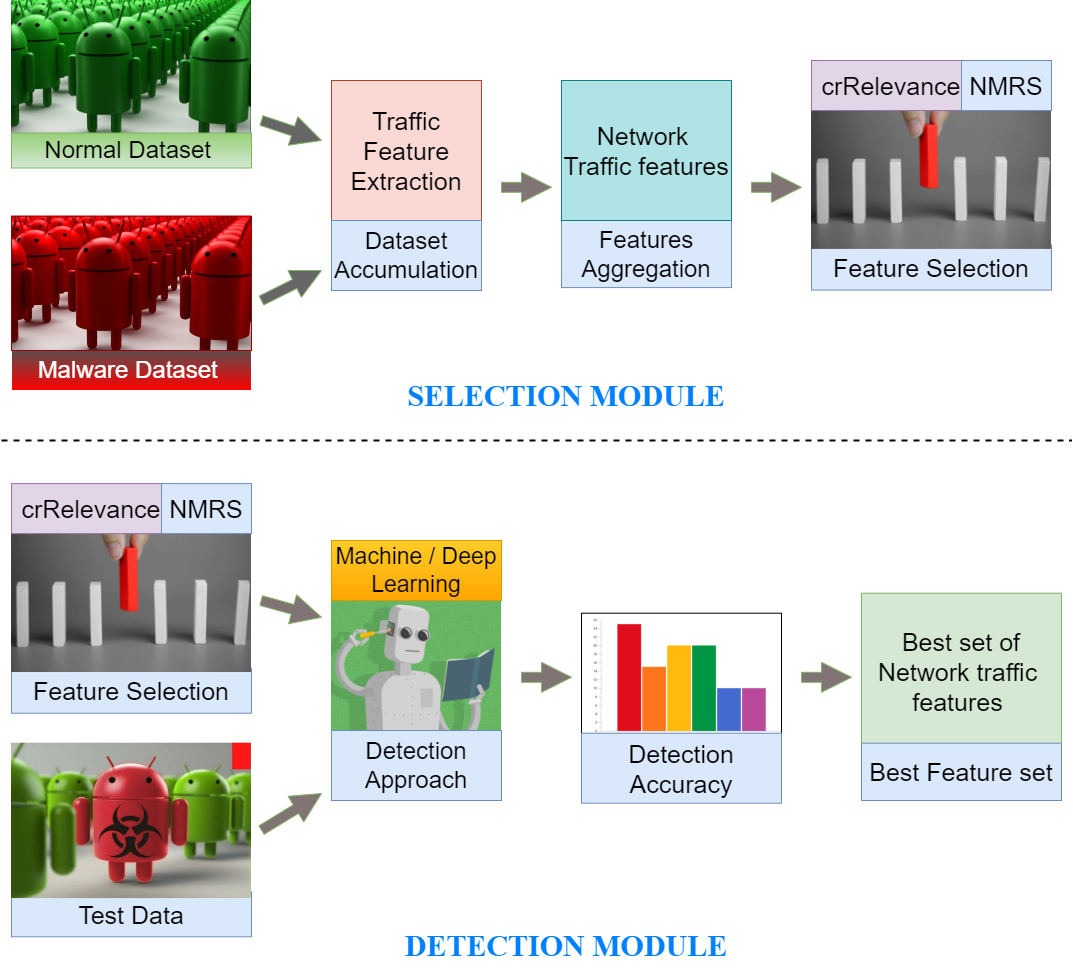}
  \caption{CorrNetDroid System Design}
  \label{Dynamic_system_design}
\end{figure}

This section delivers a comprehensive explanation of our proposed methodology. Fig. \ref{Dynamic_system_design} provides a concise overview of our proposed model \textit{CorrNetDroid}, which primarily consists of two modules. The first module, referred to as the \textit{Selection Module}, involves computing network traffic features from the TCP flows of the training dataset and ranking them using a statistical measure called \textit{crRelevance}. Such a ranking will help us grade them according to their ability to distinguish between the two class labels, normal and malware. Simultaneously, we also rank the features based on their inter-correlation score using another statistical technique called \textit{NMRS}, which would help us eliminate the redundancy between the ranked features. In the \textit{Detection Module}, we introduce a novel \textit{NMRS}-based algorithm that leverages machine learning and deep learning techniques on \textit{crRelevance} rankings to identify the optimal subset of attributes, thereby achieving higher detection accuracy. We implemented machine learning and deep learning classifiers using the Python programming language. Additionally, we utilized the Wireshark Tool to extract TCP flows from normal and malware traffic datasets. We implemented the proposed model on a desktop system featuring 8 GB of RAM, an i5-1135G7 processor, and running Windows 11 as the operating system.

The subsequent subsections provide detailed discussions on both modules of the proposed model.\\

\noindent{\textbf{SELECTION MODULE}}

\subsection{Dataset Collection}
To commence our research, we required a large dataset of mobile network traffic generated by benign and malicious applications. Hence, we obtained the network traffic data from the Canadian Institute for Cybersecurity (CIC). More specifically, we gathered four datasets, namely CICAndMal2017 \cite{CIC_And_mal2017}, CIC-InvesAndMal2019 \cite{CIC_inves2019}, CIC-AAGM2017 \cite{CIC_AAGM_2017} and USTC-TFC2016 \cite{wang2017malware}. Combining these datasets, we gathered 9,88,280 network traffic flows each for the normal and malware apps. 

\subsection{Traffic Split}

Transmission Control Protocol (TCP) flows play a crucial role in ensuring reliable and efficient communication between devices on the Internet. TCP is a connection-oriented protocol that facilitates the orderly and error-checked data delivery between applications. TCP connections start off with a three-way handshake. During this process, the sender and receiver exchange SYN (synchronization) and ACK (acknowledgment) packets to set up the connection. Moreover, different operating systems and network environments support TCP. TCP connections terminate using a four-way handshake, involving FIN (finish) and ACK flags. Its ubiquity makes it suitable for diverse applications and scenarios.
Hence, we aim to extract TCP flow-based network traffic features.

Table \ref{Network_traffic_features} summarizes the 16 traffic features used in our experiments. We extracted these features from the pcap flows of the benign and malware traffic. We have also highlighted the concise notation for each feature in the Table.  For instance, ``F\romannum{1}'' notation for ``Average\_packet\_ size'', ``F\romannum{4}" for ``Flow\_duration", etc. In the Results section, we denote the features by their notations to facilitate comprehension and interpretation of the data.

\begin{table}[t]
  \centering
\fontsize{7}{9}\selectfont
  
  \caption{List of network traffic features}
  \label{Network_traffic_features}

\begin{tabular}{|p{1.8in}|p{1.4in}|}
\hline
Average\_packet\_size   \textbf{(F\romannum{1}) }                     & Packets\_sent\_per\_second \textbf{(F\romannum{9})}      \\ \hline
Time\_interval\_between\_packets\_ sent \textbf{(F\romannum{2}) }      & Packets\_received\_per\_ second \textbf{(F\romannum{10})} \\ \hline
Time\_interval\_between\_packets\_received\textbf{(F\romannum{3})} & Packets\_received\_per\_flow \textbf{(F\romannum{11})}   \\ \hline
Flow\_duration \textbf{(F\romannum{4}) }                              & Packet\_size\_received \textbf{(F\romannum{12}) }        \\ \hline
Ratio\_of\_incoming\_to\_outgoing\_packets\textbf{(F\romannum{5})} & Bytes\_sent \textbf{(F\romannum{13})}                    \\ \hline
Ratio\_of\_incoming\_to\_outgoing\_ bytes \textbf{(F\romannum{6}) }    & Bytes\_sent\_per\_second \textbf{(F\romannum{14})}       \\ \hline
Packet\_size\_sent\textbf{ (F\romannum{7})    }                       & Bytes\_received \textbf{(F\romannum{15})}                \\ \hline
Packets\_sent\_per\_flow \textbf{(F\romannum{8}) }                    & Bytes\_received\_per\_ second \textbf{(F\romannum{16})}   \\ \hline
\end{tabular}
\end{table}

\subsection{Feature Selection}
Several techniques exist in the literature for feature selection, however, many often overlook the correlation among features. 
When selecting an optimal feature subset, it is crucial to consider both feature–feature and feature-class correlations. An ideal feature subset should consist of features highly correlated with class labels while exhibiting minimal correlation with each other \cite{hall1999correlation}. In contrast to many existing methods, we propose a feature selection technique that addresses both correlations: feature–feature and feature–class. We employ \textit{crRelevance} and \textit{NMRS} measures to achieve these correlations. In the following subsections, we discuss the working of \textit{crRelevance} and \textit{NMRS} measures. 

\subsubsection{\textit{\textbf{{crRelevance}: Feature-Class Correlation}}}

\textit{crRelevance} is a measure used to evaluate the ability of a feature to distinguish between various class labels \cite{borah2014statistical}. It yields a value within the interval of $0$ to $1$. The subsequent definitions establish the conceptual foundation for \textit{crRelevance}.

\textbf{Definition 1} For a feature $f_k$ with values $\left\{x_1, x_2 \ldots x_n\right\}$ corresponding to $n$ objects or instances in the dataset, a class range can be defined as a range $\mathrm{R}=\left[r_1, r_2\right]$ such that, $\forall x_i, x_j, r_1 \leq x_i \leq r_2, r_1 \leq x_j \leq r_2 \quad$ and $\quad \operatorname{class}_{f_k}\left(x_i\right)=$ $\operatorname{class}_{f_k}\left(x_j\right)$, where $\operatorname{class}_f(x)$ is the class associated with value $\mathrm{x}$ over feature $\mathrm{f}$. In other words, a class range $\mathrm{R}=$ $\left[r_1, r_2\right]$ over feature $\mathrm{f}$ is said to be associated with class $\mathrm{A}$ \cite{borah2014statistical}, if $\forall x, r_1 \leq x \leq r_2, \operatorname{class}_f(x)=A$.

\textbf{Definition 2} Cardinality of a class range $\mathrm{R}=\left[r_1, r_2\right]$, denoted as $\operatorname{rcard}_{\mathrm{f}}(\mathrm{R})$ \cite{borah2014statistical} is defined as the cardinality of the set $\left\{x \mid x \in V, r_1 \leq x \leq r_2\right\}$, where $V$ is the set of values of all the objects for feature $f$ .

\textbf{Definition 3} The class-cardinality of a class A, denoted as ccard(A) \cite{borah2014statistical}, is defined as the cardinality of the set $\{x \mid \operatorname{class}(x)=A\}$ .

\textbf{Definition 4} Core class range of class A denoted as ccrange(A), can be defined as the highest class range associated with the class $\mathrm{A}$ in terms of class range cardinality. A range $R_i$ \cite{borah2014statistical} associated with class $\mathrm{A}$ is called the core class range of $\mathrm{A}$ if there is no $R_j$ in $F$, such that $\operatorname{rcard}\left(R_j\right)$ $>\operatorname{rcard}\left(R_i\right)$.

Now, core range-based relevance of a class A for feature $f_i$, denoted by $\operatorname{\textit{crRelevance}}_{\mathrm{f}_{\mathrm{i}}}^{\text {class }}(\mathrm{A})$, is defined as follows.
\begin{equation}
\operatorname{\textit{crRelevance}}_{\mathrm{f}_{\mathrm{i}}}^{\text {class }}(\mathrm{A})=\frac{\operatorname{rcard}(\operatorname{ccrange}(\mathrm{A}))}{\operatorname{ccard}(\mathrm{A})}
\end{equation}

For a dataset $\mathrm{D}$, the core class relevance of a feature $f_i \in \mathrm{F}$, denoted by \textit{crRelevance} $\left(\mathrm{f}_{\mathrm{i}}\right)$ \cite{borah2014statistical}, can be defined as the highest \textit{crRelevance} for a given class $A_i$. 
\begin{equation}
\operatorname{\textit{crRelevance}}\left(f_i\right)=\max _{1 \leq j \leq n} \operatorname{\textit{crRelevance}}_{f_i}^{\text {class }}\left(A_j\right)
\end{equation}

In this work, we have two numerical-type continuous datasets comprising 16 features for two classes, benign and malware. 
We obtain the \textit{crRelevance\_normal} and \textit{crRelevane\_malware} scores for each feature after applying the above definitions to both the training datasets individually.  Further, we compute the absolute difference between the \textit{crRelevance\_normal} and \textit{crRelevane\_malware} scores for each feature to identify the most distinguishing features. Higher distinguishing features will be at the top of the ranked list irrespective of their preference for either class . We would like to point out that we considered the absolute difference instead of the normal difference to avoid the accumulation of the best features on the endpoints of the ranked list rather than on the top. The features were ranked based on their scores, making sure that the feature with the highest \textit{crRelevance score} is placed at the top, demonstrating its greater ability to differentiate across class labels.

\subsubsection{\textit{\textbf{{NMRS}- Feature-Feature Correlation}}}

The \textit{Normalized Mean Residue Similarity} (\textit{NMRS}) technique ranks features according to their relevance to the output variable. However, it does not consider redundant features. In general, closely related features often produce similar results or have a similar impact on the output data. Hence, to assess if two features exhibit similar patterns, it is essential to choose a suitable similarity measure. In other words, to find the correlation between features, we use an effective method called \textit{Normalized Mean Residue Similarity} (\textit{NMRS}). The level of concordance \cite{mahanta2012effective}  between feature  $d_1=\left(a_1, a_2, \ldots\right.$, $\left.a_n\right)$ with respect to another feature $d_2=\left(b_1, b_2, \ldots, b_n\right)$ is defined by the following Equation:
\begin{equation}
1-\frac{\sum_{i=1}^n\left|a_i-a_{\text {mean }}-b_i+b_{\text {mean }}\right|}{2 \times \max \left\{\sum_{i=1}^n\left|\left(a_i-a_{\text {mean }}\right)\right|, \sum_{i=1}^n\left|\left(b_i-b_{\text {mean }}\right)\right|\right\}},
\end{equation}
where $a_{\text {mean }}$ is the mean of all the elements of feature $d_1$;
\begin{equation}
a_{\text {mean }}=\left\{a_1+a_2+\ldots+a_n\right\} / n,
\end{equation}
and $b_{\text {mean }}$ is the mean of all the elements of feature $d_2$ ;
\begin{equation}
b_{\text {mean }}=\left\{b_1+b_2+\ldots+b_n\right\} / n \text {. }
\end{equation}

In our proposed work, because we wish to find the feature–feature correlation between all possible pairs without any relation to their classes, we concatenate the normal and malware traffic feature values. We combine the flows of both class labels for each feature and then compute the \textit{NMRS} scores. In this way, we iteratively compute the \textit{NMRS} score for each feature pair by considering them as $d_1$ and $d_2$ variables in the given \textit{NMRS} equation. 
Pearson correlation coefficient, Spearman correlation coefficient, and Mean Squared Residue are the most commonly employed similarity/correlation measurements. Nevertheless, each of these approaches has its drawbacks. Additionally, the Pearson correlation coefficient can identify not just shifting patterns but also scaling and other patterns that are often undesirable and may involve features with a substantial disparity in their expression levels. The Spearman Rank correlation coefficient uses rank values to compute correlation, however, it is unable to identify patterns including shifting or scaling. The mean squared residue is effective in identifying changing patterns, but it cannot function in a mutual mode, hence, it cannot determine the correlation between a pair of features. Therefore, we have chosen \textit{NMRS} as our preferred choice for the feature–feature correlation.

\textbf{Using the \textit{NMRS}-based proposed feature selection method that further uses \textit{crRelevance} developed rankings, we answer our research question two, i.e., how to rank features in order to identify the most distinctive and least correlated features.}
\\

\noindent\textbf{DETECTION MODULE}

We applied several machine learning and deep learning classifiers \cite{witten2002data}  in our detection approach. Primarily, we employed nine commonly utilized classifiers, including Decision Trees, Random Forest, Bagging Classifier, Gaussian Naive Bayes, Gradient Boosting, AdaBoost, Multilayer Perceptron, Dense Neural Network, and Convolutional Neural network in our experiments. We performed all the experiments with ten-fold cross-validation. In the subsequent section, we detail the results obtained from the proposed approach.

\section{Results and Discussion}

In this section, we delve into the experimental results acquired through the proposed \textit{CorrNetDroid} model. It's important to note that we utilized separate datasets for both training and testing purposes. As elaborated in Section III.A, we accumulated 9,88,280 TCP flows, each from the benign and malware categories, by compiling four datasets from various repositories. Of these, we use 6,94,261 flows of each category for training and the remaining 2,94,019 flows for testing. We named this as ``\textit{Testing Dataset}''. In the following subsections, we initially explore the rankings derived from the two statistical measures employed in our study, namely \textit{crRelevance} and \textit{NMRS}. Subsequently, we delineate the detection results obtained on the \textit{Testing Dataset}. Moreover, we conducted a comparative analysis of our proposed model with similar models designed for Android malware detection. 

\subsection{Features Ranking}

 We applied the statistical measure \textit{crRelevance} separately to the benign and malware traffic features to identify the distinguishing features. As an output, the technique produces a pair of feature-class correlation scores for each feature. Furthermore, we used the sorted absolute difference between the \textit{crRelevance}\_normal and \textit{crRelevance}\_malware of features to rank them. Table \ref{crRelevance_rankings} summarizes the network features ranked according to the absolute difference between the category scores. Table \ref{crRelevance_rankings} highlights that the feature ``Packet\_size\_received (F\romannum{12})''  is ranked at the top with the highest absolute difference between category scores. Similarly, the feature ``Bytes\_sent\_per\_second (F\romannum{14})'' has the worst absolute difference and may have the worst distinguishing characteristics trait.

%

Our proposed feature selection and detection algorithm uses both feature–class, and feature–feature correlation, i.e., \textit{crRelevance} and \textit{NMRS} rankings, to identify the best subset of features. Hence, in the next step, we applied \textit{NMRS} to all possible feature pairs (120 pairs) and ranked them in decreasing order of their correlation score. Table \ref{NMRS_rankings} summarizes the top 20 and bottom 20 traffic feature pairs, based on their \textit{NMRS} correlation scores. The table can be understood as follows. The features ``Average\_packet\_size (F\romannum{1})'' and ``Packet\_size\_received (F\romannum{12})'' have the best correlation score and thus have the best correlation between them. Similarly, the features ``Time\_interval\_between\_ packets\_sent (F\romannum{2})'' and ``Packets\_received\_per\_second (F\romannum{10})'' have the worst correlation between them.

\begin{table}[h]
  \centering
\fontsize{7}{9}\selectfont
  
  \caption{Traffic features ranked using various statistical tests }
  \label{Statistical_test_rankings}

\begin{tabular}{|p{.6in}|p{.6in}|p{0.4in}|p{.4in}|p{.4in}|} \hline 
\textbf{Chi-Square   on Normal} & \textbf{Chi-Square on Malware} & \textbf{ANOVA} & \textbf{Mann-Whitney} & \textbf{Kruskal-Wallis} \\ \hline
F\romannum{5}                              & F\romannum{5}                             & F\romannum{3}             & F\romannum{5}                    & F\romannum{3}                      \\ \hline
F\romannum{2}                              & F\romannum{6}                             & F\romannum{4}             & F\romannum{7}                    & F\romannum{5}                      \\ \hline
F\romannum{3}                              & F\romannum{7}                             & F\romannum{7}             & F\romannum{14}                   & F\romannum{12}                     \\ \hline
F\romannum{6}                              & F\romannum{1}                             & F\romannum{5}             & F\romannum{10}                   & F\romannum{12}                     \\ \hline
F\romannum{9}                              & F\romannum{2}                             & F\romannum{2}             & F\romannum{11}                   & F\romannum{6}                      \\ \hline
F\romannum{10}                             & F\romannum{12}                            & F\romannum{16}            & F\romannum{9}                    & F\romannum{8}                      \\ \hline
F\romannum{7}                              & F\romannum{10}                            & F\romannum{1}             & F\romannum{2}                    & F\romannum{15}                     \\ \hline
F\romannum{1}                              & F\romannum{4}                             & F\romannum{10}            & F\romannum{13}                   & F\romannum{7}                      \\ \hline
F\romannum{4}                              & F\romannum{8}                             & F\romannum{8}             & F\romannum{16}                   & F\romannum{4}                      \\ \hline
F\romannum{12}                             & F\romannum{11}                            & F\romannum{11}            & F\romannum{4}                    & F\romannum{14}                     \\ \hline
F\romannum{8}                              & F\romannum{3}                             & F\romannum{13}            & F\romannum{15}                   & F\romannum{10}                     \\ \hline
F\romannum{11}                             & F\romannum{9}                             & F\romannum{6}             & F\romannum{8}                    & F\romannum{16}                     \\ \hline
F\romannum{14}                             & F\romannum{13}                            & F\romannum{15}            & F\romannum{6}                    & F\romannum{11}                     \\ \hline
F\romannum{16}                             & F\romannum{16}                            & F\romannum{9}             & F\romannum{1}                    & F\romannum{13}                     \\ \hline
F\romannum{13}                             & F\romannum{15}                            & F\romannum{12}            & F\romannum{12}                   & F\romannum{9}                      \\ \hline
F\romannum{15}                             & F\romannum{14}                            & F\romannum{14}            & F\romannum{13}                   & F\romannum{2}                      \\ \hline
\end{tabular}
\end{table}

 \subsection{Comparison of \textit{crRelevance} with other statistical tests} 

In this subsection, we compare the performance of our proposed model with that of some commonly used statistical tests for Android malware detection. Our proposed approach involved computing the feature– class correlation, i.e., \textit{crRelevance} ranking, and further deploying an \textit{NMRS}-based detection algorithm to select the best set of features that produce higher detection accuracy. Hence, to compare our model’s performance, we deploy the same \textit{NMRS}-based detection algorithm on the following statistical measures: 

\begin{enumerate}

    \item \textbf{Chi-Square} - The chi-square test evaluates the disparity between expected and observed values to assess whether the observed deviations fall within an acceptable range, and is represented by the following formula:
\begin{equation}
{\chi_c}^2=\sum\frac{(O_i - E_i)^2}{E_i}
\end{equation}

where: \\
c=Degrees of freedom, \\
 O=Observed value(s), and  \\
  E=Expected value(s). \\

\item \textbf{ANOVA} - ANOVA, short for Analysis of Variance, is a statistical technique employed to examine variations between the means of different groups within a sample. It is commonly used to evaluate the null hypothesis that suggests the means of three or more groups are the same. In our case, normal and malware values of the same feature from the training dataset are taken as the two groups. The one-way ANOVA $F$-statistic is calculated using the following formula:
\begin{equation}
F=\frac{\text { Between-Group Variance (MSB) }}{\text { Within-Group Variance (MSW) }}
\end{equation}

where:\\
MSB (Mean Square Between) is the variance among the group means and MSW (Mean Square Within) is the average of the variances within each group. 

 \item \textbf{Mann-Whitney} - The Mann-Whitney U test, alternatively referred to as the Wilcoxon rank-sum test, is a statistical method that assesses whether there is a significant difference between two independent and randomly chosen groups. The $U$ statistic is calculated using the following formula:
\begin{equation}
U_1=R_1-\frac{n_1 \cdot\left(n_1+1\right)}{2} \end{equation}
\begin{equation}
U_2=R_2-\frac{n_2 \cdot\left(n_2+1\right)}{2}
\end{equation}
where:\\
$U_1$ and $U_2$ are the $U$ statistics, $R_1$ and $R_2$ are the sums of ranks and $n_1$ and $n_2$ are the sample sizes for Group 1 and Group 2, respectively.

\item \textbf{Kruskal-Wallis} - The Kruskal-Wallis test is a non-parametric test used to determine whether there are statistically significant differences between three or more independent groups.  The test assesses whether the samples originate from the same distribution or if at least one of the samples is different from the others. The formula for the test statistic is as follows :
\begin{equation}
H=\frac{12}{N(N+1)} \sum_{i=1}^k \frac{R_i^2}{n_i}-3(N+1)
\end{equation}
where:\\
$N$ is the total number of observations across all groups,\\
$k$ is the number of groups,\\
$R_i$ is the sum of ranks for group $i$, and \\
$n_i$ is the number of observations in group $i$. \\

\end{enumerate}

Table \ref{Statistical_test_rankings} summarizes the individual test rankings when we apply chi-square on the normal feature set and chi-square on the malware feature set along with other measures such as ANOVA, Mann–Whitney U test, and Kruskal– Wallis test. As can be seen from the table, ``Ratio\_of\_incoming\_to\_ outgoing\_packets (F\romannum{5})'' has been ranked as the most distinguishing feature by Mann– Whitney as well as both rankings of the chi-square test, whereas, according to ANOVA and Kruskal– Wallis, the feature named ``Time\_interval\_between\_ packets\_received (F\romannum{3})'' resides at the top of the table as the best feature. 

For comparison, we ranked network traffic features using the four statistical tests described above and further applied our proposed \textit{NMRS}-based algorithm on the \textit{\textit{Testing Dataset}} to determine their corresponding accuracies. Table \ref{D1_Statistical_tests_accuracy} summarizes the results and it can be understood as follows. When we apply the \textit{NMRS}-based proposed approach on chi-square rankings (on benign training dataset) and Kruskal– Wallis rankings, we achieve the highest detection accuracy of 96.22\% after eliminating 14 features out of the total lot of 16 traffic features. With two features, namely ``Ratio\_of\_incoming\_to\_ outgoing\_packets (F\romannum{5})'' and ``Time\_interval\_between\_ packets\_received (F\romannum{3})'', the highest detection accuracy can be achieved. At the same time, the highest detection accuracy of 96.37\% can be achieved upon considering two features, namely ``Ratio\_of\_incoming\_to\_ outgoing\_packets (F\romannum{5})'' and ``Time\_interval\_between\_ packets\_sent (F\romannum{2})'' when we apply the \textit{NMRS}-based proposed algorithm on Mann– Whitney test rankings and chi-square test rankings for malware dataset. Similarly, when we apply the \textit{NMRS}-based proposed algorithm to ANOVA test rankings, we obtain the highest detection accuracy of 98.10\% while considering two features, namely ``Time\_interval\_between\_ packets\_received (F\romannum{3})'' and ``Flow\_Duration (F\romannum{4})''. When we apply the \textit{NMRS}-based proposed algorithm to the \textit{crRelevance} rankings used in our work, we obtain an accuracy of 99.5\% with two features, namely ``Packet\_size\_received (F\romannum{12})'' and ``Time\_interval\_between\_ packets\_received (F\romannum{3})''. Hence, our model outperforms other similar statistical tests when we apply our proposed \textit{NMRS} algorithm to their developed rankings.  

\subsection{Comparsion with other related works}

In this subsection, we assess the performance of our proposed model with another similar work incorporating TCP flows. We do so by implementing the approach followed by the authors in \cite{shabtai2014mobile}. Similar to the working of their proposed model, we chose Decision Table and REPTree classifiers as the base learners for the \textit{local learning model}. We trained these classifiers with all 16 features. According to the authors, when a benign feature vector was evaluated against the models developed during the learning phase, there is a greater likelihood that the projected value would closely resemble or match the observed value. If the forecasts deviate significantly from the actual values of the relevant attributes, it indicates a higher probability that the observed vector originates from a distinct distribution. Therefore, we ranked the features in order of their probability of deviation from the pattern of benign traffic features. Table \ref{Mobile_malware_paper} summarizes the traffic features ranked in order of their deviation from normal traffic behavior. As depicted by the table, the feature ``Packets\_sent\_per\_flow (F\romannum{8})'' shows the highest probability of coming from an abnormal event. At the same time, the feature ``Ratio\_of\_incoming\_to\_ outgoing\_packets (F\romannum{5})'' scored the lowest related to the least probability of coming from an abnormal event.

After ranking the network traffic features in order of their probability of deviation from benign traffic patterns, we proceed to find the best feature set with higher detection accuracy. Table \ref{Mobile_malware_accuracy} summarizes the detection results using Decision Table and RepTree classifiers as base learners. As it can be seen, upon eliminating four features out of the total set of 16 features, we managed to achieve the highest detection accuracy of 95.85\% while using both Decision Table and RepTree classifiers as base learners, In other words, on considering the top 12 features, namely ``F\romannum{8}, F\romannum{15}, F\romannum{11}, F\romannum{4}, F\romannum{13}, F\romannum{9}, F\romannum{14}, F\romannum{2}, F\romannum{16}, F\romannum{3}, F\romannum{10}, and  F\romannum{6}", both the classifiers display the same detection accuracy of 95.85\%. Whereas, with our \textit{NMRS}-based proposed algorithm applied to \textit{crRelevance} rankings, we could achieve a higher detection accuracy of 99.5\%. Hence, we can conclude that our proposed approach outperforms this similar network traffic based Android malware detection model.

Furthermore, we evaluated our work with other resembling works of Android malware detection. Table \ref{Comparision of work} summarizes this comparison. Some studies have attempted to determine the probability of deviation from normal traffic features’ patterns, whereas others have ranked the features using tests such as chi-square, information gain, and frequency ranges to select the best subset of features. As the table summarizes, our proposed model outperforms all these studies with respect to detection accuracy. Therefore, we can deduce that our suggested model surpasses numerous state-of-the-art methodologies in the existing body of knowledge for detecting Android malware.  

\subsection{Limitations}

Next, we outline few limitations of the proposed methodology. The proposed model ranks TCP-based network traffic features for detection; hence, it falls under the category of dynamic analysis. The path of dynamic analysis overcomes several limitations of static analysis but also poses some barriers. Not all Android malware samples generate network traffic. It has been observed that certain types of malware may transmit text messages discreetly in the background, without producing any noticeable network traffic. Hence, network traffic-based detection mechanisms cannot detect such samples. Dynamic analysis tools can introduce performance overhead as they monitor and analyze the execution of the program. This overhead may affect the timing and behavior of the software, potentially masking certain performance-related issues. In addition, certain mobile assaults may occur as a result of colluding apps, when harmful behavior is spread throughout multiple apps rather than being confined to a single app. Nevertheless, the current version of the proposed paradigm does not specifically address the issue of colluding apps. Hence, in our future endeavors, we intend to focus on identifying colluding applications in order to improve the detection capability of the suggested model.

\section{Conclusion and Future Work}
In this study, we ranked the network traffic features in order of their correlation with the class and amongst themselves using two statistical measures, namely \textit{crRelevance} and \textit{Normalized Mean Residue Similarity} (\textit{NMRS}). Subsequently, we proposed a novel \textit{NMRS}-based detection algorithm to select the best and inversely correlated features by applying various machine learning and deep learning techniques. The experimental results highlight that our proposed \textit{NMRS}-based detection algorithm on \textit{crRelevance} rankings can effectively reduce the feature set while detecting Android malware with 99.50\% accuracy on considering two network traffic features, namely ``Packet\_size\_received'' and ``Time\_interval\_between\_ packets\_received''. Furthermore, our results showed that our proposed method is better than other statistical tests such as Chi-Square, ANOVA, Mann–Whitney U test and Kruskal–Wallis test. Moreover, the proposed model can detect Android malware with better accuracy than various state-of-the-art techniques. In our future work, we aim to enhance the capabilities of our model by including malware category and family classification along with the binary classification performed in this study. We also aim to integrate static features to deal with the limitations of dynamic analysis and possibly build a robust hybrid detection model. 

\vskip -2\baselineskip plus -1fil

\bibliographystyle{IEEEtran}
\bibliography{Manuscript}

\begin{IEEEbiography}[{\includegraphics[width=1in,height=1.4in,clip,keepaspectratio]{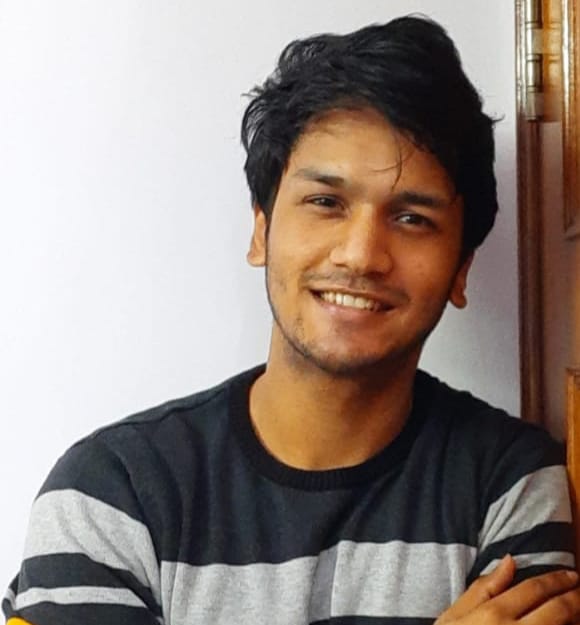}}]{Yash Sharma} is currently a doctoral student specializing in Android Security in the Department of Applied Mathematics at Delhi Technological University, Delhi, India. He holds a postgraduate degree in Mathematics from the same institution and has conducted research in Network Security as part of his academic work. He has published papers in the field of Android Security in SCIE journals, including the Journal of Network and Computer Applications (Elsevier),  Multimedia Tools and Applications, International Journal of Information Security (Springer).

\end{IEEEbiography}

\vskip -2\baselineskip plus -1fil

\begin{IEEEbiography}[{\includegraphics[width=1in,height=1.4in,clip,keepaspectratio]{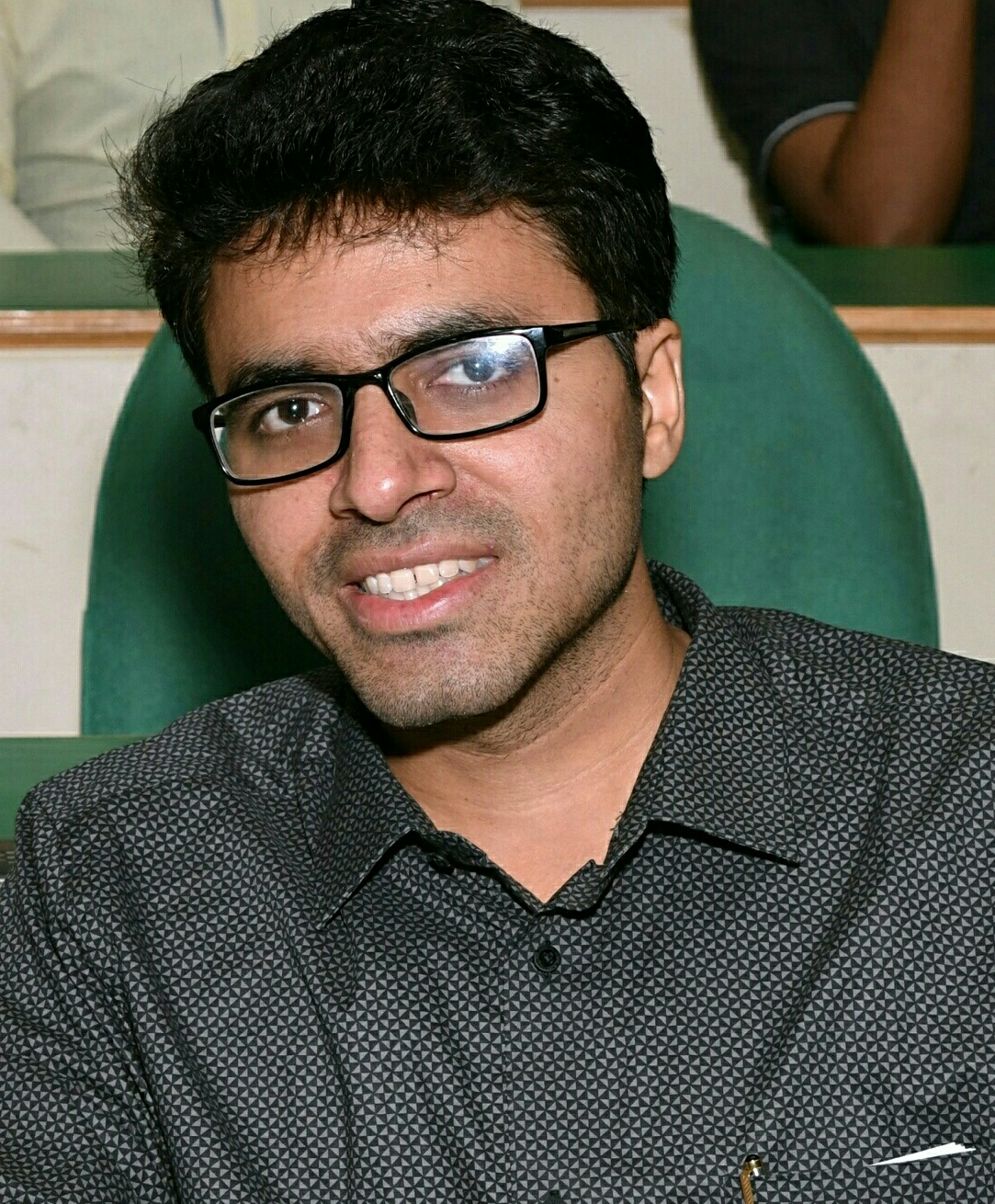}}]{Anshul Arora}
is currently working as an Assistant Professor in the Discipline of Mathematics and Computing, Department of Applied Mathematics, Delhi Technological University Delhi, India. He has done his Ph.D. from the Department of Computer Science and Engineering, Indian Institute of Technology Roorkee, India. He has published several research papers in his field of expertise which are Mobile Security, Mobile Malware Detection, Network Traffic Analysis, and Blockchain. He is the reviewer of several renowned journals such as IEEE Transactions on Information Forensics and Security, IEEE Transactions on Computational Social Systems, Expert Systems with Applications, Computers \& Security, etc.


\end{IEEEbiography}

\end{document}